\documentclass{IEEEtran}
\usepackage{ifpdf}
\usepackage[nospace]{cite}
\usepackage{graphicx}
\usepackage{color}
\input{epsf.sty}

\usepackage{bm}
\usepackage[cmex10]{amsmath}
\usepackage{amssymb}
\usepackage{array}
\usepackage{syntonly}
\usepackage{amsthm}
\usepackage{amsfonts}
\usepackage{epsfig}
\usepackage{latexsym}

\newcommand{\Define}{\stackrel{\triangle}{=}}

\DeclareMathOperator{\Tr}{Tr}

\usepackage{fixltx2e}
\usepackage{url}

\theoremstyle{definition} 
\theoremstyle{definition} 
\theoremstyle{definition} 

\begin{document}
\title{{\LARGE 
MIMO DF Relay Beamforming for Secrecy with Artificial Noise, Imperfect CSI, 
and Finite-Alphabet Input\thanks{This work was supported in part by the 
Indo-French Centre for Applied Mathematics.}} }
\author{
{\large Sanjay Vishwakarma and A. Chockalingam} \\
Email: sanjay@ece.iisc.ernet.in, achockal@ece.iisc.ernet.in \\
{\normalsize Department of ECE, Indian Institute of Science, 
Bangalore 560012}
}

\maketitle

\begin{abstract}
In this paper, we consider decode-and-forward (DF) relay beamforming
with imperfect channel state information (CSI), cooperative artificial
noise (AN) injection, and finite-alphabet input in the presence of an 
user and $J$ non-colluding eavesdroppers. The communication between the 
source and the user is aided by a multiple-input-multiple-output (MIMO) 
DF relay. We use the fact that a wiretap code consists of two parts: 
$i)$ common message (non-secret), and $ii)$ secret message. The source 
transmits two independent messages: $i)$ common message (non-secret),
and $ii)$ secret message. The common message is transmitted at a fixed 
rate $R_{0}$, and it is intended for the user. The secret message is 
also intended for the user but it should be kept secret from the $J$ 
eavesdroppers. The source and the MIMO DF relay operate under individual 
power constraints. In order to improve the secrecy rate, the MIMO relay 
also injects artificial noise. The CSI on all the links are assumed to
be imperfect and CSI errors are assumed to be norm bounded. In order to 
maximize the worst case secrecy rate, we maximize the worst case link 
information rate to the user subject to: $i)$ the individual power 
constraints on the source and the MIMO relay, and $ii)$ the best case 
link information rates to $J$ eavesdroppers be less than or equal to 
$R_{0}$ in order to support a fixed common message rate $R_{0}$.
Numerical results showing the effect of perfect/imperfect CSI, 
presence/absence of AN with finite-alphabet input on the secrecy rate 
are presented.
\end{abstract}
{\em keywords:}
{\em {\footnotesize
MIMO relay beamforming, physical layer security, multiple eavesdroppers, 
artificial noise, imperfect CSI, finite-alphabet input, semi-definite 
programming.
}} 

\section{Introduction}
\label{sec_mimo_df_relay_bf_jamming_finitealphabet_1}
The physically degraded discrete memoryless wiretap channel model 
considered by Wyner in \cite{ir1} opened the path for reliable and 
secure information transmission using physical layer techniques. 
Subsequent extension to discrete memoryless broadcast channel and 
Gaussian channel was done in \cite{ir2} and \cite{ir3}, respectively. 
A wireless network can be easily eavesdropped due to the broadcast 
nature of wireless transmission. However, using physical layer techniques
(e.g. wiretap codes, beamforming using multiple antennas, artificial noise 
injection etc.), a wireless network can be secured from getting eavesdropped. 
Achievable secrecy rate and capacity in single and multiple antenna wiretap 
channels have been reported by many authors, e.g., 
\cite{ir4, ir5, ir6, ir7, ir8, ir9}. 

A relay, operating in decode-and-forward (DF) or amplify-and-forward (AF) 
mode, can act as an intermediate node and help improving the secrecy rate 
\cite{ir10}. DF and AF relay beamforming techniques for secrecy under 
perfect/imperfect channel state information (CSI) have been well studied 
in the literature, e.g. \cite{ir11, ir12, ir13, ir14, ir15, ir16}.  
In these works, the transmit codeword symbols belong to an infinite 
constellation (Gaussian). However, in a practical communication system, 
the codeword symbols will belong to a finite alphabet set, e.g., $M$-ary 
alphabets. The effect of finite constellation on secrecy rate has been 
reported in \cite{ir17, ir18, ir19, ir20, irx, ir21, ir22}. In \cite{ir21}, 
DF relay beamforming for secrecy with finite alphabet has been considered.
There it was shown that when the source power and relay beamforming vector 
obtained for Gaussian alphabet, when used with finite alphabet, could 
lead to zero secrecy rate. A power control algorithm was suggested to 
alleviate the loss in secrecy rate. Motivated by the above works, in 
this paper, we consider secrecy rate in DF relay beamforming with 
finite-alphabet input using a MIMO relay. The considered system consists 
of a source node, a destination node, and multiple non-colluding 
eavesdroppers. A DF MIMO relay aids the communication between the source 
and destination. It is known that secrecy rate can be improved through 
the use of artificial noise (AN) injection 
\cite{ir8}, \cite{ir11}, \cite{irz}, \cite{ir26}, \cite{ir27}, \cite{ir16}. 
In this work, we allow the MIMO relay to inject AN in addition to relaying 
the information symbol from the source. Consequently, we solve for both the 
optimum source power, signal beamforming weights as well as 
the AN covariance matrix at the MIMO relay. Since 
the CSI will not be perfect in practice, we consider a norm-bounded CSI error 
model and investigate the effect of imperfect CSI on the secrecy rate. We use 
the fact that a wiretap code consists of two parts: $i)$ common message 
(non-secret), and $ii)$ secret message. The source transmits two independent 
messages: $i)$ common message (non-secret), and $ii)$ secret message. The 
common message is transmitted at a fixed rate $R_{0}$, and its intended 
for the destination node. The secret message is also intended for the 
destination node but it should be kept secret from $J$ eavesdroppers. 
The source and the MIMO DF relay operate under individual power 
constraints. In order to maximize the worst case secrecy rate, we 
maximize the worst case link information rate to the user subject to: 
$i)$ the individual power constraints on the source and the MIMO DF 
relay, and $ii)$ the best case link information rates to $J$ eavesdroppers 
be less than or equal to $R_{0}$ in order to support a fixed common message 
rate $R_{0}$. Numerical results showing the effect of perfect/imperfect CSI, 
presence/absence of AN with finite-alphabet input on the secrecy rate are 
presented. 

$\bf{Notations:}$ $\boldsymbol{A} \in 
\mathbb{C}^{N_{1} \times N_{2}}$ implies that $\boldsymbol{A}$ is 
a complex matrix of dimension $N_{1} \times N_{2}$. 
$\boldsymbol{A} \succeq \boldsymbol{0}$ and 
$\boldsymbol{A} \succ \boldsymbol{0}$ imply
that $\boldsymbol{A}$ is a positive semidefinite matrix and
positive definite matrix, respectively.
Identity matrix is denoted by $\boldsymbol{I}$.
Transpose and complex conjugate 
transpose operations are denoted by $[.]^{T}$ and $[.]^{\ast}$, 
respectively.  $\mathbb{E}[.]$ denotes expectation operator. 
$\parallel\hspace{-1mm}.\hspace{-1mm}\parallel$ denotes 2-norm operator.
Trace of matrix $\boldsymbol{A} \in \mathbb{C}^{N \times N}$ is denoted 
by $\Tr(\boldsymbol{A})$.
$\boldsymbol{\psi}_{} \in \mathbb{C}^{N_{} \times 1} \sim \mathcal{CN}(\boldsymbol{0}, \boldsymbol{\Psi}_{})$ 
implies that $\boldsymbol{\psi}_{}$ is a circularly symmetric complex 
Gaussian random vector with mean vector $\boldsymbol{0}$ and covariance 
matrix $\boldsymbol{\Psi}_{}$.

\section{System model}
\label{sec_mimo_df_relay_bf_jamming_finitealphabet_2}
\begin{figure}
\center
\includegraphics[totalheight=6.5cm,width=6.5cm]{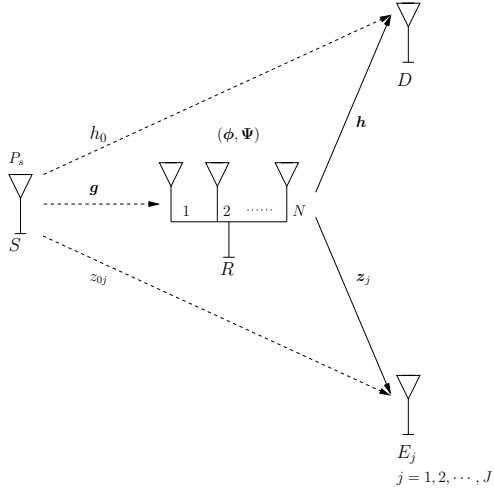}
\caption{System model for MIMO DF relaying.} 
\label{fig_mimo_df_relay_bf_jamming_finitealphabet_1}
\end{figure}

Consider a DF cooperative relaying scheme which consists of a source 
node $S$ having single transmit antenna, a MIMO DF relay node $R$ having 
$N$ receive/transmit antennas, a destination node $D$ having single receive 
antenna, and $J$ non-colluding eavesdropper nodes $E_{1},E_{2},\cdots,E_{J}$ 
having single receive antenna each. The system model is shown in Fig. 
\ref{fig_mimo_df_relay_bf_jamming_finitealphabet_1}. In addition to the 
links from relay to destination node and relay to eavesdropper nodes, we 
assume direct links from source to destination node and source to 
eavesdropper nodes. The complex channel gain vector between the source 
and the relay is denoted by 
$\boldsymbol{g}=[g_{1},g_{2},\cdots,g_{N}]^{T} \in {\mathbb{C}}^{N\times 1}$. 
Likewise, the channel gain vector between the relay and the destination 
node $D$ is denoted by  
$\boldsymbol{h}=[h_{1},h_{2},\cdots,h_{N}] \in {\mathbb{C}}^{1\times N}$, 
and the channel gain vector between the relay and the $j$th eavesdropper 
node $E_{j}$, $1 \leq j \leq J$, is denoted by  
$\boldsymbol{z}_{j}=[z_{1j},z_{2j},\cdots,z_{Nj}]\in {\mathbb{C}}^{1\times N}$. 
The channel gains on the direct links from the source to $D$ and 
the source to $E_{j}$ are denoted by $h_{0}$ and $z_{0j}$, 
respectively.

The MIMO relay operates in half duplex mode, and the communication 
happens in two hops. Each hop is divided into $n$ channel uses. 
We use the fact that a wiretap code consists of two parts: 
$i)$ common message (non-secret), and $ii)$ secret message. 
In the first hop of transmission, the source $S$ transmits two 
independent messages $W_{0}$ and $W_{1}$ which are equiprobable over 
$\{1,2,\cdots,2^{2nR_{0}}\}$ and $\{1,2,\cdots,2^{2nR_{s}(R_{0})}\}$, 
respectively. $W_{0}$ is the common message which is transmitted at a 
fixed rate $R_{0}$ and its intended for the destination $D$. $W_{1}$ 
is a secret message which is transmitted at some rate $R_{s}(R_{0})$ 
and its also intended only for $D$ and it should be kept secret from 
all $E_{j}$s. For each $W_{0}$ and $W_{1}$ drawn independently and 
equiprobably from the sets $\{1,2,\cdots,2^{2nR_{0}}\}$ and 
$\{1,2,\cdots,2^{2nR_{s}(R_{0})}\}$, respectively,
the source $S$ maps $W_{0}$ and $W_{1}$ to a codeword 
$\{ x_{m} \}^{n}_{m = 1}$ of length $n$. Each symbol, $x_{m}$, in the 
codeword is independent and equiprobable over a complex finite-alphabet 
set $\mathbb{A} = \{ a_{1}, a_{2}, \cdots, a_{M} \} $ of size $M$ with
${\mathbb{E}} [ {x_{m}} ] = 0$, and ${\mathbb{E}} [ {|x_{m}|}^{2} ] = 1$.
The source is constrained by the available power $P_{S}$ and it transmits 
the weighted symbol which is $\sqrt{P_{s}}x_{m}$ in the $m$th channel use, 
where $1 \leq m \leq n$, and $0 \leq P_{s} \leq P_{S}$. Hereafter, we will 
denote the symbol $x_{m}$ of the codeword $\{ x_{m} \}^{n}_{m = 1}$ by $x$, 
and we will consider only one channel use. 

Let $\boldsymbol{y}_{R}$, ${y}_{D}$, and ${y}_{E_{1j}}$ denote the 
received signals at the MIMO relay $R$, destination $D$, and $j$th 
eavesdropper $E_{j}$, respectively, in the first hop. We have
\begin{eqnarray}
\boldsymbol{y}_{R} \ &=& \ \sqrt{P_{s}}\boldsymbol{g} x \ + \ \boldsymbol{\eta}_{R}, 
\label{eqn_mimo_df_relay_bf_jamming_finitealphabet_1} \\  
y_{D_{1}} \ &=& \ \sqrt{P_{s}}h_{0}          x \ +  \ \eta_{D_{1}},         
\label{eqn_mimo_df_relay_bf_jamming_finitealphabet_2} \\
y_{E_{1j}} \ &=& \ \sqrt{P_{s}}z_{0j}          x \ +  \ \eta_{E_{1j}},         
\label{eqn_mimo_df_relay_bf_jamming_finitealphabet_3}
\end{eqnarray}
where $\boldsymbol{\eta}_{R} ( \sim {\mathcal{CN}}(\boldsymbol{0}, N_{0}\boldsymbol{I}))$,
${\eta}_{D_{1}} ( \sim {\mathcal{CN}}(0, N_{0}) )$, and
${\eta}_{E_{1j}} ( \sim {\mathcal{CN}}(0, N_{0}) )$ are receiver
noise components and are assumed to be independent.  

In the second hop of transmission, MIMO relay applies the complex weight 
$\boldsymbol{\phi} = [\phi_{1},\phi_{2},\cdots,\phi_{N}]^{T} \in {\mathbb{C}}^{N \times 1}$ 
on the successfully decoded symbol $x$ and retransmits it.
In order to improve the secrecy rate, MIMO relay also injects the 
artificial noise
$\boldsymbol{\psi} \in {\mathbb{C}}^{N \times 1}(\sim {\mathcal{CN}}(\boldsymbol{0}, \boldsymbol{\Psi}))$. 
The symbol transmitted by the MIMO relay on the $i$th, $1 \leq i \leq N$, 
antenna is $\phi_{i} x + \psi_{i}$. Let ${y}_{D_{2}}$, and ${y}_{E_{2j}}$ 
denote the received signals at the destination $D$, and $j$th eavesdropper 
$E_{j}$, respectively, in the second hop. We have
\begin{eqnarray}
y_{D_{2}} \ &=& \ \boldsymbol{h} \boldsymbol{\phi}x \ + \ \boldsymbol{h} \boldsymbol{\psi} \ + \ \eta_{D_{2}}, 
\label{eqn_mimo_df_relay_bf_jamming_finitealphabet_4} \\
y_{E_{2j}} \ &=& \ \boldsymbol{z}_{j} \boldsymbol{\phi}x \ + \ \boldsymbol{z}_{j} \boldsymbol{\psi} \ + \ \eta_{E_{2j}}, 
\label{eqn_mimo_df_relay_bf_jamming_finitealphabet_5}
\end{eqnarray}
where 
${\eta}_{D_{2}} ( \sim {\mathcal{CN}}(0, N_{0}) )$, and
${\eta}_{E_{2j}} ( \sim {\mathcal{CN}}(0, N_{0}) )$ are receiver
noise components and are assumed to be independent. Using 
(\ref{eqn_mimo_df_relay_bf_jamming_finitealphabet_2}), 
(\ref{eqn_mimo_df_relay_bf_jamming_finitealphabet_4}), and
(\ref{eqn_mimo_df_relay_bf_jamming_finitealphabet_3}), 
(\ref{eqn_mimo_df_relay_bf_jamming_finitealphabet_5}),
we rewrite the received signals at $D$ and $E_{j}$ 
in the following vector forms, respectively:
\begin{eqnarray}
\boldsymbol{y}_{D} \ &=& \ [y_{D_{1}}, \ y_{D_{2}}]^{T} \nonumber \\
&=& \ {[ \sqrt{P_{s}}h_{0}, \ \boldsymbol{h}\boldsymbol{\phi}]}^{T}x \ + \  {[\eta_{D_{1}}, \ \boldsymbol{h}\boldsymbol{\psi} + \eta_{D_{2}}]}^{T},
\label{eqn_mimo_df_relay_bf_jamming_finitealphabet_6} \\
\boldsymbol{y}_{E_{j}} \ &=& \ [y_{E_{1j}}, \ y_{E_{2j}}]^{T} \nonumber \\ 
&=& \ {[ \sqrt{P_{s}}z_{0j}, \ \boldsymbol{z}_{j}\boldsymbol{\phi}]}^{T}x \ + \  {[\eta_{E_{1j}}, \ \boldsymbol{z}_{j}\boldsymbol{\psi} + \eta_{E_{2j}}]}^{T}.
\label{eqn_mimo_df_relay_bf_jamming_finitealphabet_7}
\end{eqnarray}
We assume that the MIMO relay's transmit power, denoted by $P_{r}$,
is constrained by the available power $P_{R}$. This implies that 
\begin{eqnarray}
P_{r} \ &=& \ {\mathbb{E}}\{{\parallel \hspace{-1mm} (\boldsymbol{\phi}x + \boldsymbol{\psi}) \hspace{-1mm} \parallel}^{2} \} \nonumber \\
      \ &=& \ {\parallel \hspace{-1mm} \boldsymbol{\phi} \hspace{-1mm} \parallel}^{2} \  +  \  \Tr(\boldsymbol{\Psi}) \ \leq \ P_{R}.
\label{eqn_mimo_df_relay_bf_jamming_finitealphabet_8}
\end{eqnarray}
We also assume that the channel remains static over the entire codeword 
transmit duration. Further, denoting the secret message decoded at the 
MIMO relay $R$ and destination $D$ by $\widehat{W}^{\footnotesize{R}}_{1}$
and $\widehat{W}^{\footnotesize{D}}_{1}$, respectively, the reliability 
constraints at $R$ and $D$ and the perfect secrecy constraints at $E_{j}$s are as 
follows:
\begin{eqnarray}
\text{Pr}(\widehat{W}^{\footnotesize{R}}_{1} \neq W_{1}) &\leq& \epsilon_{n},  
\quad 
\text{Pr}(\widehat{W}^{\footnotesize{D}}_{1} \neq W_{1}) \ \ \leq \ \ \epsilon_{n},  \nonumber \\
\frac{1}{2n}I(W_{1}; \boldsymbol{y}^{2n}_{E_{j}} )        &\leq& \epsilon_{n}, \ \ \forall j \ = \ 1,2,\cdots,J,   \nonumber
\end{eqnarray}
where $\boldsymbol{y}^{2n}_{E_{j}}$ is
the received signal vector at $E_{j}$ in $2n$ channel uses, and 
$\epsilon_{n} \rightarrow 0$ as $n \rightarrow \infty$. 
We also note that the reliability constraints at the MIMO relay $R$ and
destination $D$ for the secret message also 
ensure the reliability of the common message.

\section{DF relay beamforming - perfect CSI}
\label{sec_mimo_df_relay_bf_jamming_finitealphabet_3}
In this section, we assume that the CSI on all the links are known perfectly.
Using (\ref{eqn_mimo_df_relay_bf_jamming_finitealphabet_1}), 
(\ref{eqn_mimo_df_relay_bf_jamming_finitealphabet_6}), and 
(\ref{eqn_mimo_df_relay_bf_jamming_finitealphabet_7}), 
we get the $S$-$R$, $S$-$D$, and $S$-$E_{j}$ link information rates, 
respectively, as follows:
\begin{eqnarray}
\frac{1}{2}I(x; \boldsymbol{y}_{R}) &=& \frac{1}{2}I \bigg( \frac{ P_{s} {\parallel \hspace{-1mm} \boldsymbol{g} \hspace{-1mm} \parallel}^{2} }{ N_{0} }\bigg), 
\label{eqn_mimo_df_relay_bf_jamming_finitealphabet_9} \\
\frac{1}{2}I(x; \boldsymbol{y}_{D} ) &=& \frac{1}{2} I \bigg( \frac{ P_{s} { | h_{0} | }^{2} }{ N_{0} } + \frac{ \boldsymbol{h} \boldsymbol{\phi} \boldsymbol{\phi}^{\ast} \boldsymbol{h}^{\ast} }{ N_{0} + \boldsymbol{h} \boldsymbol{\Psi} \boldsymbol{h}^{\ast} } \bigg), 
\label{eqn_mimo_df_relay_bf_jamming_finitealphabet_10} \\
\frac{1}{2}I(x; \boldsymbol{y}_{E_{j}} ) &=& \frac{1}{2} I \bigg(  \frac{ P_{s} { | z_{0j} | }^{2} }{ N_{0} } + \frac{ \boldsymbol{z}_{j} \boldsymbol{\phi} \boldsymbol{\phi}^{\ast} \boldsymbol{z}^{\ast}_{j} }{ N_{0} + \boldsymbol{z}_{j} \boldsymbol{\Psi} \boldsymbol{z}^{\ast}_{j} } \bigg), 
\label{eqn_mimo_df_relay_bf_jamming_finitealphabet_11}
\end{eqnarray}
where 
\begin{eqnarray}
I(\rho) \ \Define \  \frac{1}{M} \sum\limits^{M}_{l = 1} \int p_{n}\big(y_{_{}} - \sqrt{\rho} a_{l}\big) \nonumber \\
\log_{2} \frac{p_{n}(y_{_{}} - \sqrt{\rho} a_{l})}{\frac{1}{M} \sum\limits^{M}_{m = 1}p_{n}(y_{_{}} - \sqrt{\rho} a_{m})} d y_{_{}}, 
\label{eqn_mimo_df_relay_bf_jamming_finitealphabet_12}
\end{eqnarray}
and $p_n(\theta) = \frac{1}{\pi} e^{{{-\mid \theta \mid}^{2}}}$. 
The factor $1/2$ in 
(\ref{eqn_mimo_df_relay_bf_jamming_finitealphabet_9}),
(\ref{eqn_mimo_df_relay_bf_jamming_finitealphabet_10}), and
(\ref{eqn_mimo_df_relay_bf_jamming_finitealphabet_11})
is due to two hops. Further, the MIMO relay $R$ will be able to decode 
the symbol $x$ if the following condition holds true 
\cite{ir11, ir15, ir16, ir21}:
\begin{eqnarray}
\frac{1}{2}I(x;\boldsymbol{y}_{R}) \ \geq \ \frac{1}{2}I(x;\boldsymbol{y}_{D}).
\label{eqn_mimo_df_relay_bf_jamming_finitealphabet_13}
\end{eqnarray}
In order to find the maximum achievable secrecy rate $R_{s}(R_{0})$ 
which also supports the fixed common message rate $R_{0}$, we maximize 
the $S-D$ link information rate subject to $i)$ $S-E_{j}$, $1\leq j \leq J$, 
link information rates be less than or equal to $R_{0}$, $ii)$ the 
information rate constraint in 
(\ref{eqn_mimo_df_relay_bf_jamming_finitealphabet_13}), and
$iii)$  the power constraints. The optimization problem is as follows:
\begin{eqnarray}
R_{D}(R_{0}) \ = \ \max_{ P_{s}, \ \boldsymbol{\phi},  \ \boldsymbol{\Psi} }  \ \frac{1}{2}I(x; \boldsymbol{y}_{D}) 
\label{eqn_mimo_df_relay_bf_jamming_finitealphabet_14} \\
\text{s.t.} \quad \quad  \frac{1}{2}I(x; \boldsymbol{y}_{E_j}) \ \leq \ R_{0}, \quad \forall j=1,2,\cdots,J, 
\label{eqn_mimo_df_relay_bf_jamming_finitealphabet_15} \\
\frac{1}{2}I(x;\boldsymbol{y}_{R} ) \ \geq \ \frac{1}{2}I(x;\boldsymbol{y}_{D}),
\label{eqn_mimo_df_relay_bf_jamming_finitealphabet_16} \\
0 \leq P_{s} \leq P_{S}, \quad \boldsymbol{\Psi} \succeq \boldsymbol{0}, \quad
{\parallel \hspace{-1mm} \boldsymbol{\phi} \hspace{-1mm} \parallel}^{2} \  +  \  \Tr(\boldsymbol{\Psi}) \ \leq \ P_{R}.
\label{eqn_mimo_df_relay_bf_jamming_finitealphabet_17}
\end{eqnarray}
Having obtained $R_{D}(R_{0})$ from 
(\ref{eqn_mimo_df_relay_bf_jamming_finitealphabet_14}),
the maximum achievable secrecy rate $R_{s}(R_{0})$ for a given
common message rate $R_{0}$ is \cite{ir9}
\begin{eqnarray}
R_{s}(R_{0}) \ = \ { \{ R_{D}(R_{0}) - R_{0} \} }^{+}, 
\label{eqn_mimo_df_relay_bf_jamming_finitealphabet_18}
\end{eqnarray}
where $ { \{ \alpha \} }^{+} = \max(0, \alpha)$.
From the constraint in (\ref{eqn_mimo_df_relay_bf_jamming_finitealphabet_16}), 
its obvious that the upper bound for $S-D$ link information rate, denoted by 
$R_{D}$, can be obtained by evaluating 
(\ref{eqn_mimo_df_relay_bf_jamming_finitealphabet_9}) at $P_{s} = P_{S}$.
For the values of $R_{0}$ over the interval $[0, R_{D}]$, the maximum 
achievable secrecy rate, denoted by $R_{s}$, is obtained as follows:
\begin{eqnarray}
R_{s} \ &=& \ \max_{0 \ \leq \ R_{0} \ \leq \ R_{D}} \ { \{ R_{D}(R_{0}) - R_{0} \} }^{+} 
\label{eqn_mimo_df_relay_bf_jamming_finitealphabet_19} \\
\ &=& \ \max_{0 \ \leq \ l     \ \leq \ L} \ { \{ R_{D}(l\Delta_{1}) - l\Delta_{1} \} }^{+},    
\label{eqn_mimo_df_relay_bf_jamming_finitealphabet_20}
\end{eqnarray}
where $L$ is a large positive integer, $\Delta_{1} = R_{D}/L$, $l$ is 
an integer, and $R_{0} = l \Delta_{1}$.

We solve the optimization problem 
(\ref{eqn_mimo_df_relay_bf_jamming_finitealphabet_14}) 
for a fixed $P_{s} = k \Delta_{2}$, 
where $\Delta_{2} = P_{S}/K$, $K$ is a large positive integer, and 
$1 \leq k \leq K$. Hereafter, we will assume that $P_{s}$ is known.
Further, it is shown in \cite{ir23, ir24} that for various $M$-ary 
alphabets, mutual information expression in 
(\ref{eqn_mimo_df_relay_bf_jamming_finitealphabet_12}) is a strictly 
increasing concave function in SNR. With this fact, we rewrite the
optimization problem (\ref{eqn_mimo_df_relay_bf_jamming_finitealphabet_14}) 
into the following equivalent form:
\begin{eqnarray}
\max_{\boldsymbol{\Phi}, \ \boldsymbol{\Psi}}  \ \bigg( a + \frac{ \boldsymbol{h} \boldsymbol{\Phi} \boldsymbol{h}^{\ast} }{ N_{0} + \boldsymbol{h} \boldsymbol{\Psi} \boldsymbol{h}^{\ast}  } \bigg)
\label{eqn_mimo_df_relay_bf_jamming_finitealphabet_21} 
\end{eqnarray}
\begin{eqnarray}
\text{s.t.} \quad \quad  \quad \forall j=1,2,\cdots,J, \nonumber \\
\bigg( b_{j} +  \frac{ \boldsymbol{z}_{j} \boldsymbol{\Phi} \boldsymbol{z}^{\ast}_{j} }{ N_{0} + \boldsymbol{z}_{j} \boldsymbol{\Psi} \boldsymbol{z}^{\ast}_{j} } \bigg) \ \leq \ I^{-1}(2R_{0}),  
\label{eqn_mimo_df_relay_bf_jamming_finitealphabet_22} \\
c \ \geq \ \bigg( a + \frac{ \boldsymbol{h} \boldsymbol{\Phi} \boldsymbol{h}^{\ast} }{ N_{0} + \boldsymbol{h} \boldsymbol{\Psi} \boldsymbol{h}^{\ast}  } \bigg) 
\label{eqn_mimo_df_relay_bf_jamming_finitealphabet_23} \\
\boldsymbol{\Phi} \succeq \boldsymbol{0}, \ rank(\boldsymbol{\Phi}) = 1, \ \boldsymbol{\Psi} \succeq \boldsymbol{0}, \
\Tr( \boldsymbol{\Phi} +  \boldsymbol{\Psi}) \ \leq \ P_{R},
\label{eqn_mimo_df_relay_bf_jamming_finitealphabet_24}
\end{eqnarray}
where $\boldsymbol{\Phi} = \boldsymbol{\phi} \boldsymbol{\phi}^{\ast}$,
$a = \big( \frac{P_{s} {|h_{0}|}^{2}}{N_{0}} \big)$, 
$b_{j} = \big( \frac{P_{s} {|z_{0j}|}^{2}}{N_{0}} \big)$, and 
$c = \big( \frac{ P_{s} {\parallel  \boldsymbol{g} \parallel}^{2}}{N_{0}}\big)$.
Further, relaxing the $ rank(\boldsymbol{\Phi}) = 1$ constraint, 
we rewrite the above optimization problem into the following form:
\begin{eqnarray}
\max_{t, \ \boldsymbol{\Phi}, \ \boldsymbol{\Psi} }  \quad t
\label{eqn_mimo_df_relay_bf_jamming_finitealphabet_25} \\
\text{s.t.} \quad \quad
(t - a) \big( N_{0} + \boldsymbol{h} \boldsymbol{\Psi} \boldsymbol{h}^{\ast} \big) - (\boldsymbol{h} \boldsymbol{\Phi} \boldsymbol{h}^{\ast}) \ \leq \ 0, 
\label{eqn_mimo_df_relay_bf_jamming_finitealphabet_59} \\
\forall j \ = \ 1,2,\cdots,J, \quad \nonumber \\ \big( \boldsymbol{z}_{j} \boldsymbol{\Phi} \boldsymbol{z}^{\ast}_{j} \big) - \Big( I^{-1}(2R_{0}) - b_{j} \Big) ( N_{0} + \boldsymbol{z}_{j} \boldsymbol{\Psi} \boldsymbol{z}^{\ast}_{j}  ) \ \leq \ 0,  
\label{eqn_mimo_df_relay_bf_jamming_finitealphabet_60} \\
(\boldsymbol{h} \boldsymbol{\Phi} \boldsymbol{h}^{\ast}) - (c - a) \big( N_{0} + \boldsymbol{h} \boldsymbol{\Psi} \boldsymbol{h}^{\ast} \big)  \ \leq \ 0, 
\label{eqn_mimo_df_relay_bf_jamming_finitealphabet_61} \\
\boldsymbol{\Phi} \succeq \boldsymbol{0}, \quad \boldsymbol{\Psi} \succeq \boldsymbol{0}, \quad
\Tr( \boldsymbol{\Phi}  +  \boldsymbol{\Psi}) \ \leq \ P_{R}.
\label{eqn_mimo_df_relay_bf_jamming_finitealphabet_26}
\end{eqnarray}
The above problem can be easily solved using bisection method \cite{ir25}. 
The initial search interval in the bisection method can be taken as 
$[0, \ c]$. In the appendix, we show that the solution $\boldsymbol{\Phi}$ 
of the above problem has rank 1. Further, denoting the maximum value of 
$t$ by $t_{max}$, the secrecy rate is obtained as follows:
\begin{eqnarray}
R_{s}(R_{0}) \ = \ { \bigg \{ \frac{1}{2} I(t_{max}) - R_{0} \bigg \} }^{+}.
\label{eqn_mimo_df_relay_bf_jamming_finitealphabet_27}
\end{eqnarray}

\section{DF relay beamforming - imperfect CSI}
\label{sec_mimo_df_relay_bf_jamming_finitealphabet_4}
In this section, we assume that each receiver has perfect knowledge 
of its CSI. We also assume that the control unit which computes the 
source power, signal beamforming vector and AN covariance matrix has imperfect CSI on all links. 
The imperfection in CSI is modeled as follows \cite{ir15, ir26, ir27}:
\begin{eqnarray}
\boldsymbol{g}=\widehat{\boldsymbol{g}}+\boldsymbol{e}_{\boldsymbol{g}}, \quad 
h_{0} = \widehat{h}_{0} + e_{ h_{0} }, \quad
\boldsymbol{h} = \widehat{ \boldsymbol{h} } + \boldsymbol{e}_{ \boldsymbol{h} }, \nonumber \\
\forall j = 1,2,\cdots,J, \quad 
z_{0j} = \widehat{z}_{0j} + e_{ z_{0j} }, \quad
\boldsymbol{z}_{j} = \widehat{ \boldsymbol{z} }_{j} + \boldsymbol{e}_{ \boldsymbol{z}_{j} }, 
\label{eqn_mimo_df_relay_bf_jamming_finitealphabet_28} 
\end{eqnarray}
where 
$\widehat{ \boldsymbol{g} }$, 
$\widehat{h}_{0}$,
$\widehat{ \boldsymbol{h} }$, 
$\widehat{z}_{0j}$,
$\widehat{ \boldsymbol{z} }_{j}$ 
are the available CSI estimates, and
$\boldsymbol{e}_{ \boldsymbol{g} }$,
$e_{ h_{0} }$,
$\boldsymbol{e}_{ \boldsymbol{h} }$,
$e_{ z_{0j} }$, 
$\boldsymbol{e}_{ \boldsymbol{z}_{j} }$
are the corresponding CSI errors. 
We assume that the CSI errors are bounded, i.e.,
\begin{eqnarray}
\parallel \hspace{-1mm} \boldsymbol{e}_{ \boldsymbol{g} } \hspace{-1mm} \parallel \leq \epsilon_{ \boldsymbol{g} }, \quad
| e_{ h_{0} } | \leq \epsilon_{ h_{0} }, \quad
\parallel \hspace{-1mm} \boldsymbol{e}_{ \boldsymbol{h} } \hspace{-1mm} \parallel \leq \epsilon_{ \boldsymbol{h} }, \nonumber \\
\forall j=1,2,\cdots,J, \quad | e_{ z_{0j} } | \leq \epsilon_{ z_{0j} }, \quad
\parallel \hspace{-1mm} \boldsymbol{e}_{ \boldsymbol{z}_{j} } \hspace{-1mm} \parallel \leq \epsilon_{ \boldsymbol{z}_{j} }.
\label{eqn_mimo_df_relay_bf_jamming_finitealphabet_29} 
\end{eqnarray}
With the above CSI error model, we write the rank relaxed optimization problem 
(\ref{eqn_mimo_df_relay_bf_jamming_finitealphabet_21}) as follows:
\begin{eqnarray}
\max_{ \boldsymbol{\Phi}, \ \boldsymbol{\Psi} } \ \min_{ \boldsymbol{e}_{ \boldsymbol{h} } }  \
\bigg( \frac{ aN_{0} + (\widehat{ \boldsymbol{h} } + \boldsymbol{e}_{ \boldsymbol{h} }) \big( a \boldsymbol{\Psi} + \boldsymbol{\Phi} \big) (\widehat{ \boldsymbol{h} } + \boldsymbol{e}_{ \boldsymbol{h} })^{\ast} }{ N_{0} + (\widehat{ \boldsymbol{h} } + \boldsymbol{e}_{ \boldsymbol{h} })  \boldsymbol{\Psi} (\widehat{ \boldsymbol{h} } + \boldsymbol{e}_{ \boldsymbol{h} })^{\ast} } \bigg) 
\label{eqn_mimo_df_relay_bf_jamming_finitealphabet_30} 
\end{eqnarray}

{\small
\begin{eqnarray}
\text{s.t.} \quad \quad {\parallel \hspace{-1mm} \boldsymbol{e}_{ \boldsymbol{h} } \hspace{-1mm} \parallel}^{2} \leq \epsilon^{2}_{ \boldsymbol{h} }, 
\label{eqn_mimo_df_relay_bf_jamming_finitealphabet_31}  \\
\left \{
\begin{array}{cc}
\forall j=1,2,\cdots,J, \\
\mathop{\max } \limits_{ \boldsymbol{e}_{ \boldsymbol{z}_{j} } } \ \bigg( \frac{ b_{j}N_{0} + (\widehat{ \boldsymbol{z} }_{j} + \boldsymbol{e}_{ \boldsymbol{z}_{j} }) \big( b_{j} \boldsymbol{\Psi} + \boldsymbol{\Phi} \big) (\widehat{ \boldsymbol{z} }_{j} + \boldsymbol{e}_{ \boldsymbol{z}_{j} })^{\ast} }{ N_{0} + (\widehat{ \boldsymbol{z} }_{j} + \boldsymbol{e}_{ \boldsymbol{z}_{j} })  \boldsymbol{\Psi} (\widehat{ \boldsymbol{z} }_{j} + \boldsymbol{e}_{ \boldsymbol{z}_{j} })^{\ast} } \bigg) \\ \leq \ I^{-1}(2R_{0}), \\ 
\text{s.t.} \quad \quad {\parallel \hspace{-1mm} \boldsymbol{e}_{ \boldsymbol{z}_{j} } \hspace{-1mm} \parallel}^{2} \leq \epsilon^{2}_{ \boldsymbol{z}_{j} }, 
\end{array}
\right \}
\label{eqn_mimo_df_relay_bf_jamming_finitealphabet_32}  \\
\left \{
\begin{array}{cc}
c \ \geq \
\mathop{\max } \limits_{ \boldsymbol{e}_{ \boldsymbol{h} } } \Big(a_{max} +  \frac{ (\widehat{ \boldsymbol{h} } + \boldsymbol{e}_{ \boldsymbol{h} }) \boldsymbol{\Phi} (\widehat{ \boldsymbol{h} } + \boldsymbol{e}_{ \boldsymbol{h} })^{\ast} }{N_{0} + (\widehat{ \boldsymbol{h} } + \boldsymbol{e}_{ \boldsymbol{h} }) \boldsymbol{\Psi} (\widehat{ \boldsymbol{h} } + \boldsymbol{e}_{ \boldsymbol{h} })^{\ast}} \Big) \\
\text{s.t.} \quad \quad {\parallel \hspace{-1mm} \boldsymbol{e}_{ \boldsymbol{h} } \hspace{-1mm} \parallel}^{2} \leq \epsilon^{2}_{ \boldsymbol{h} },
\end{array}
\right \}
\label{eqn_mimo_df_relay_bf_jamming_finitealphabet_33} \\
\boldsymbol{\Phi} \succeq \boldsymbol{0}, \quad \boldsymbol{\Psi} \succeq \boldsymbol{0}, \quad \Tr(\boldsymbol{\Phi} + \boldsymbol{\Psi}) \ \leq \ P_R, 
\label{eqn_mimo_df_relay_bf_jamming_finitealphabet_34}
\end{eqnarray}
}

\vspace{-4mm}
where
\begin{eqnarray}
a_{} = \Big( \frac{ P_{s} {| |\widehat{h}_{0}| - \epsilon_{h_{0}} |}^{2} }{ N_{0} } \Big) \quad \text{if} \quad ( | \widehat{h}_{0} | > \epsilon_{h_{0}} ), \quad 0 \quad \text{else}, 
\label{eqn_mimo_df_relay_bf_jamming_finitealphabet_35} \\
b_{j} = \Big( \frac{ P_{s} {| |\widehat{z}_{0j}| + \epsilon_{z_{0j}} |}^{2} }{ N_{0} } \Big) , 
\label{eqn_mimo_df_relay_bf_jamming_finitealphabet_36} \\
c     = \Big( \frac{ P_{s} {| \parallel \hspace{-1mm} \widehat{ \boldsymbol{g} } \hspace{-1mm} \parallel - \epsilon_{ \boldsymbol{g} } |}^{2} } { N_{0} } \Big) \quad \text{if} \quad ( \parallel \hspace{-1mm} \widehat{ \boldsymbol{g} } \hspace{-1mm} \parallel > \epsilon_{ \boldsymbol{g} } ), \quad 0 \quad \text{else},
\label{eqn_mimo_df_relay_bf_jamming_finitealphabet_37} \\
a_{max} = \Big( \frac{ P_{s} { | |\widehat{h}_{0}| + \epsilon_{ h_{0} } | }^{2} }{ N_{0} } \Big).
\label{eqn_mimo_df_relay_bf_jamming_finitealphabet_38}
\end{eqnarray}
The objective function in 
(\ref{eqn_mimo_df_relay_bf_jamming_finitealphabet_30}) corresponds 
to the worst case $S-D$ link information rate over the region of 
CSI error uncertainty. The constraint in 
(\ref{eqn_mimo_df_relay_bf_jamming_finitealphabet_32}) corresponds 
to the best case $S-E_{j}$ link information rate over the region of 
CSI error uncertainty. The constraint in 
(\ref{eqn_mimo_df_relay_bf_jamming_finitealphabet_33}) is associated 
with the information rate constraint in 
(\ref{eqn_mimo_df_relay_bf_jamming_finitealphabet_23}), i.e.,
the worst case information rate to the MIMO relay $R$ over the 
region of CSI error uncertainty should be greater than or equal to 
the best case information rate to destination $D$.

Solving the optimization problem 
(\ref{eqn_mimo_df_relay_bf_jamming_finitealphabet_30}) is hard due 
to the presence of $ \boldsymbol{e}_{ \boldsymbol{h}}$ in both 
the numerator and denominator of the objective function in 
(\ref{eqn_mimo_df_relay_bf_jamming_finitealphabet_30}) and the 
constraint in (\ref{eqn_mimo_df_relay_bf_jamming_finitealphabet_33}). 
Similarly, $ \boldsymbol{e}_{ \boldsymbol{z}_{j} } $ appears in both 
the numerator and denominator of the constraint in 
(\ref{eqn_mimo_df_relay_bf_jamming_finitealphabet_32}). 
So, by independently constraining the various quadratic
terms appearing in the objective function in 
(\ref{eqn_mimo_df_relay_bf_jamming_finitealphabet_30}) 
and the constraints in 
(\ref{eqn_mimo_df_relay_bf_jamming_finitealphabet_32}),
(\ref{eqn_mimo_df_relay_bf_jamming_finitealphabet_33}),
we get the following lower bound for the above
optimization problem:
\begin{eqnarray}
\max_{ \boldsymbol{\Phi}, \ \boldsymbol{\Psi}, \atop{r_{1}, \ r_{2}, \ r_{3}, \ r_{4}, \atop{s_{1j}, \ s_{2j}, \ j=1,2,\cdots,J} } } \ \ \frac{ r_{1} }{ r_{2} }
\label{eqn_mimo_df_relay_bf_jamming_finitealphabet_39} 
\end{eqnarray}
\begin{eqnarray}
\text{s.t.} \quad \quad \boldsymbol{\Phi} \succeq \boldsymbol{0}, \quad \boldsymbol{\Psi} \succeq \boldsymbol{0}, \quad \Tr(\boldsymbol{\Phi} + \boldsymbol{\Psi}) \ \leq \ P_R, 
\label{eqn_mimo_df_relay_bf_jamming_finitealphabet_40} \\
\forall \boldsymbol{e}_{ \boldsymbol{h} }\quad \text{s.t.} \quad {\parallel \hspace{-1mm} \boldsymbol{e}_{ \boldsymbol{h} } \hspace{-1mm} \parallel}^{2} \leq {\epsilon}^{2}_{ \boldsymbol{h}  } \ \Longrightarrow \ \nonumber \\  
0 \ \leq \ r_{1} \ \leq \ aN_{0} + (\widehat{ \boldsymbol{h} } + \boldsymbol{e}_{ \boldsymbol{h} }) \big( a \boldsymbol{\Psi} + \boldsymbol{\Phi} \big) (\widehat{ \boldsymbol{h} } + \boldsymbol{e}_{ \boldsymbol{h} })^{\ast},
\label{eqn_mimo_df_relay_bf_jamming_finitealphabet_41} \\
\forall \boldsymbol{e}_{ \boldsymbol{h} }\quad \text{s.t.} \quad {\parallel \hspace{-1mm} \boldsymbol{e}_{ \boldsymbol{h} } \hspace{-1mm} \parallel}^{2} \leq {\epsilon}^{2}_{ \boldsymbol{h}  } \ \Longrightarrow \ \nonumber \\  
N_{0} + (\widehat{ \boldsymbol{h} } + \boldsymbol{e}_{ \boldsymbol{h} })  \boldsymbol{\Psi} (\widehat{ \boldsymbol{h} } + \boldsymbol{e}_{ \boldsymbol{h} })^{\ast} \ \leq \ r_{2},
\label{eqn_mimo_df_relay_bf_jamming_finitealphabet_42} \\
\frac{s_{1j}} { s_{2j} } \ \leq \ I^{-1} (2R_{0}), \quad \forall j=1,2,\cdots,J
\label{eqn_mimo_df_relay_bf_jamming_finitealphabet_43} \\
\forall \boldsymbol{e}_{ \boldsymbol{z}_{j} }\quad \text{s.t.} \quad {\parallel \hspace{-1mm} \boldsymbol{e}_{ \boldsymbol{z}_{j} } \hspace{-1mm} \parallel}^{2} \leq {\epsilon}^{2}_{ \boldsymbol{z}_{j}  } \ \Longrightarrow \ \nonumber \\  
b_{j}N_{0} + (\widehat{ \boldsymbol{z}_{j} } + \boldsymbol{e}_{ \boldsymbol{z}_{j} }) \big( b_{j} \boldsymbol{\Psi} + \boldsymbol{\Phi} \big) (\widehat{ \boldsymbol{z}_{j} } + \boldsymbol{e}_{ \boldsymbol{z}_{j} })^{\ast} \ \leq \ s_{1j},
\label{eqn_mimo_df_relay_bf_jamming_finitealphabet_44} \\
\forall \boldsymbol{e}_{ \boldsymbol{z}_{j} }\quad \text{s.t.} \quad {\parallel \hspace{-1mm} \boldsymbol{e}_{ \boldsymbol{z}_{j} } \hspace{-1mm} \parallel}^{2} \leq {\epsilon}^{2}_{ \boldsymbol{z}_{j}  } \ \Longrightarrow \ \nonumber \\  
0 \ \leq \ s_{2j} \ \leq \  N_{0} + (\widehat{ \boldsymbol{z} }_{j} + \boldsymbol{e}_{ \boldsymbol{z}_{j} })  \boldsymbol{\Psi} (\widehat{ \boldsymbol{z} }_{j} + \boldsymbol{e}_{ \boldsymbol{z}_{j} })^{\ast},
\label{eqn_mimo_df_relay_bf_jamming_finitealphabet_45} 
\end{eqnarray}
\begin{eqnarray}
c \ \geq \ \Big( a_{max} + \frac{r_{3}}{r_{4}} \Big), 
\label{eqn_mimo_df_relay_bf_jamming_finitealphabet_46} \\
\forall \boldsymbol{e}_{ \boldsymbol{h} }\quad \text{s.t.} \quad {\parallel \hspace{-1mm} \boldsymbol{e}_{ \boldsymbol{h} } \hspace{-1mm} \parallel}^{2} \leq {\epsilon}^{2}_{ \boldsymbol{h}  } \ \Longrightarrow \ \nonumber \\  
(\widehat{ \boldsymbol{h} } + \boldsymbol{e}_{ \boldsymbol{h} })  \boldsymbol{\Phi} (\widehat{ \boldsymbol{h} } + \boldsymbol{e}_{ \boldsymbol{h} })^{\ast} \ \leq \ r_{3},
\label{eqn_mimo_df_relay_bf_jamming_finitealphabet_47} \\
\forall \boldsymbol{e}_{ \boldsymbol{h} }\quad \text{s.t.} \quad {\parallel \hspace{-1mm} \boldsymbol{e}_{ \boldsymbol{h} } \hspace{-1mm} \parallel}^{2} \leq {\epsilon}^{2}_{ \boldsymbol{h}  } \ \Longrightarrow \ \nonumber \\  
0 \ \leq \ r_{4} \ \leq \ N_{0} + (\widehat{ \boldsymbol{h} } + \boldsymbol{e}_{ \boldsymbol{h} }) \boldsymbol{\Psi} (\widehat{ \boldsymbol{h} } + \boldsymbol{e}_{ \boldsymbol{h} })^{\ast}.
\label{eqn_mimo_df_relay_bf_jamming_finitealphabet_48}
\end{eqnarray}
The quadratic inequality constraints in 
(\ref{eqn_mimo_df_relay_bf_jamming_finitealphabet_41}) and 
(\ref{eqn_mimo_df_relay_bf_jamming_finitealphabet_42}) are 
associated with the objective function in 
(\ref{eqn_mimo_df_relay_bf_jamming_finitealphabet_30}). The 
constraint in (\ref{eqn_mimo_df_relay_bf_jamming_finitealphabet_43}), 
and the quadratic inequality constraints in 
(\ref{eqn_mimo_df_relay_bf_jamming_finitealphabet_44}) and 
(\ref{eqn_mimo_df_relay_bf_jamming_finitealphabet_45}) are associated with
the constraint in (\ref{eqn_mimo_df_relay_bf_jamming_finitealphabet_32}).
Similarly, the constraint in 
(\ref{eqn_mimo_df_relay_bf_jamming_finitealphabet_46}), and 
the quadratic inequality constraints in 
(\ref{eqn_mimo_df_relay_bf_jamming_finitealphabet_47}) and 
(\ref{eqn_mimo_df_relay_bf_jamming_finitealphabet_48}) are associated with
the constraint in (\ref{eqn_mimo_df_relay_bf_jamming_finitealphabet_33}).

Further, using $S$-procedure \cite{ir25}, we transform the quadratic inequality 
constraints in 
(\ref{eqn_mimo_df_relay_bf_jamming_finitealphabet_41}),
(\ref{eqn_mimo_df_relay_bf_jamming_finitealphabet_42}),
(\ref{eqn_mimo_df_relay_bf_jamming_finitealphabet_44}),
(\ref{eqn_mimo_df_relay_bf_jamming_finitealphabet_45}),
(\ref{eqn_mimo_df_relay_bf_jamming_finitealphabet_47}), and
(\ref{eqn_mimo_df_relay_bf_jamming_finitealphabet_48}),
into the following linear matrix inequality (LMI) forms, respectively:

{\scriptsize
\begin{eqnarray}
&\hspace{-100mm} & \hspace{-45mm} r_{1} \geq 0, \ \ \lambda_{1} \geq 0, 
\boldsymbol{A}_{1} \Define \nonumber \\
\left [
\begin{array}{cc}
\big( a \boldsymbol{\Psi} + \boldsymbol{\Phi} \big) + \lambda_{1} \boldsymbol{I} & \big( a \boldsymbol{\Psi} + \boldsymbol{\Phi} \big) \widehat{ \boldsymbol{h} }^{\ast} \\ \widehat{ \boldsymbol{h} }^{} \big( a \boldsymbol{\Psi} + \boldsymbol{\Phi} \big)^{\ast} & aN_{0} + \widehat{ \boldsymbol{h} }^{} \big( a \boldsymbol{\Psi} + \boldsymbol{\Phi} \big)\widehat{ \boldsymbol{h} }^{\ast} -r_{1} - \lambda_{1}\epsilon^{2}_{ \boldsymbol{h} } 
\end{array} \right ], \nonumber \\ 
&\hspace{-100mm} &\hspace{-35mm} \lambda_{2} \geq 0, \ \ \boldsymbol{A}_{2} \Define \nonumber \\
\left [ 
\begin{array}{cc}
-\boldsymbol{\Psi} + \lambda_{2} \boldsymbol{I} & -\boldsymbol{\Psi} \widehat{ \boldsymbol{h} }^{\ast} \\ -\widehat{ \boldsymbol{h} }^{} \boldsymbol{\Psi}^{\ast} & -N_{0} - \widehat{ \boldsymbol{h} }^{} \boldsymbol{\Psi}\widehat{ \boldsymbol{h} }^{\ast} + r_{2} - \lambda_{2}\epsilon^{2}_{ \boldsymbol{h} } 
\end{array} \right ], \nonumber \\
&\hspace{-100mm} &\hspace{-38mm} \mu_{1j} \geq 0, \ \ \boldsymbol{B}_{1j} \Define \nonumber \\ 
\left [ 
\begin{array}{cc}
-\big( b_{j} \boldsymbol{\Psi} + \boldsymbol{\Phi} \big) + \mu_{1j} \boldsymbol{I} & -\big( b_{j} \boldsymbol{\Psi} + \boldsymbol{\Phi} \big) \widehat{ \boldsymbol{z} }^{\ast}_{j} \\ -\widehat{ \boldsymbol{z} }^{}_{j} \big( b_{j} \boldsymbol{\Psi} + \boldsymbol{\Phi} \big)^{\ast} &  -b_{j}N_{0} - \widehat{ \boldsymbol{z} }^{}_{j} \big( b_{j} \boldsymbol{\Psi} + \boldsymbol{\Phi} \big)\widehat{ \boldsymbol{z} }^{\ast}_{j} + s_{1j} - \mu_{1j}\epsilon^{2}_{ \boldsymbol{z}_{j} } 
\end{array} \right ], \nonumber \\
&\hspace{-100mm} &\hspace{-50mm} s_{2j} \geq 0, \ \ \mu_{2j} \geq 0, \ \ \boldsymbol{B}_{2j} \Define \nonumber \\ 
\left [ 
\begin{array}{cc}
\boldsymbol{\Psi} + \mu_{2j} \boldsymbol{I} & \boldsymbol{\Psi} \widehat{ \boldsymbol{z} }^{\ast}_{j} \\ \widehat{ \boldsymbol{z} }^{}_{j} \boldsymbol{\Psi}^{\ast} & N_{0} + \widehat{ \boldsymbol{z} }^{}_{j} \boldsymbol{\Psi}\widehat{ \boldsymbol{z} }^{\ast}_{j} - s_{2j} - \mu_{2j}\epsilon^{2}_{ \boldsymbol{z}_{j} } 
\end{array} \right ], \nonumber \\
&\hspace{-100mm} &\hspace{-35mm} \lambda_{3} \geq 0, \quad \boldsymbol{A}_{3} \Define \nonumber \\
\left [ 
\begin{array}{cc}
-\boldsymbol{\Phi} + \lambda_{3} \boldsymbol{I} & -\boldsymbol{\Phi} \widehat{ \boldsymbol{h} }^{\ast} \\ -\widehat{ \boldsymbol{h} }^{} \boldsymbol{\Phi}^{\ast} & - \widehat{ \boldsymbol{h} }^{} \boldsymbol{\Phi}\widehat{ \boldsymbol{h} }^{\ast} + r_{3} - \lambda_{3}\epsilon^{2}_{ \boldsymbol{h} } 
\end{array} \right ], \nonumber \\
&\hspace{-100mm} &\hspace{-45mm} r_{4} \geq 0, \quad \lambda_{4} \geq 0, \quad \boldsymbol{A}_{4} \Define \nonumber \\ 
\left[
\begin{array}{cc}
\boldsymbol{\Psi} + \lambda_{4} \boldsymbol{I} & \boldsymbol{\Psi} \widehat{ \boldsymbol{h} }^{\ast} \\ \widehat{ \boldsymbol{h} }^{} \boldsymbol{\Psi}^{\ast} & N_{0} + \widehat{ \boldsymbol{h} }^{} \boldsymbol{\Psi}\widehat{ \boldsymbol{h} }^{\ast} - r_{4} - \lambda_{4}\epsilon^{2}_{ \boldsymbol{h} } 
\end{array} \right], \nonumber 
\end{eqnarray}
}

\vspace{-4mm}
\hspace{-4.5mm} where 
$\boldsymbol{A}_{1} \succeq \boldsymbol{0}$,
$\boldsymbol{A}_{2} \succeq \boldsymbol{0}$,
$\boldsymbol{A}_{3} \succeq \boldsymbol{0}$,
$\boldsymbol{A}_{4} \succeq \boldsymbol{0}$,
$\boldsymbol{B}_{1j} \succeq \boldsymbol{0}$,
$\boldsymbol{B}_{2j} \succeq \boldsymbol{0}$.
We substitute the above LMI constraints in the optimization problem 
(\ref{eqn_mimo_df_relay_bf_jamming_finitealphabet_39}). 
We get the following equivalent form for the optimization problem 
(\ref{eqn_mimo_df_relay_bf_jamming_finitealphabet_39}):

\vspace{-1mm}
{\small
\begin{eqnarray}
\max_{ \boldsymbol{\Phi}, \ \boldsymbol{\Psi}, \atop{r_{1}, \cdots r_{4}, \ \lambda_{1},\cdots,\lambda_{4}, \atop{s_{1j}, \ s_{2j}, \ \mu_{1j}, \ \mu_{2j}, \ j=1,2,\cdots,J, \atop{r} } }  }  \ \ r
\label{eqn_mimo_df_relay_bf_jamming_finitealphabet_55} \\
\text{s.t.} \quad \quad \boldsymbol{\Phi} \succeq \boldsymbol{0}, \quad \boldsymbol{\Psi} \succeq \boldsymbol{0}, \quad \Tr(\boldsymbol{\Phi} + \boldsymbol{\Psi}) \ \leq \ P_R, & & \nonumber \\
r r_{2} - r_{1} \leq 0, \quad \forall j=1,2,\cdots,J, \quad  s_{1j} - s_{2j} I^{-1} (2R_{0}) \leq 0, & & \nonumber \\
r_{1} \geq 0, \ r_{4} \geq 0, \ \lambda_{1} \geq 0, \ \lambda_{2} \geq 0, \ \lambda_{3} \geq 0, \ \lambda_{4} \geq 0, & & \nonumber \\
\boldsymbol{A}_{1} \succeq \boldsymbol{0}, \ \boldsymbol{A}_{2} \succeq \boldsymbol{0}, \ \boldsymbol{A}_{3} \succeq \boldsymbol{0}, \ \boldsymbol{A}_{4} \succeq \boldsymbol{0}, & & \nonumber \\
s_{2j} \geq 0, \ \mu_{1j} \geq 0, \ \mu_{2j} \geq 0, \ \boldsymbol{B}_{1j} \succeq \boldsymbol{0},\ \boldsymbol{B}_{2j} \succeq \boldsymbol{0}, & & \nonumber \\
r_{3} - (c - a_{max}) r_{4} \leq 0.
\label{eqn_mimo_df_relay_bf_jamming_finitealphabet_56}
\end{eqnarray}
}

\vspace{-1mm}
\hspace{-4.5mm}
The above problem can be solved using the bisection method as discussed 
in Section \ref{sec_mimo_df_relay_bf_jamming_finitealphabet_3}. The 
initial search interval in the bisection method can be taken as 
$[0, \ c]$, where $c$ is as defined in 
(\ref{eqn_mimo_df_relay_bf_jamming_finitealphabet_37}).
Further, denoting the maximum value of $r$ by $r_{max}$, the lower bound 
on the secrecy rate is obtained as follows:
\begin{eqnarray}
R_{s}(R_{0}) \ \geq \ { \bigg \{ \frac{1}{2} I(r_{max}) - R_{0} \bigg \} }^{+}.
\label{eqn_mimo_df_relay_bf_jamming_finitealphabet_57}
\end{eqnarray} 

\section{Results and discussions}
\label{sec_mimo_df_relay_bf_jamming_finitealphabet_5}
In this section, we present numerical results on the secrecy rate for 
BPSK alphabet (i.e., $M =2$), with/without AN, perfect/imperfect CSI 
conditions. We assume that $N = 2$, $J = 1,2,3$, $N_{0} = 1$, 
$P_{s} = 0$ dB, and $P_{R} = 9$ dB.

\begin{figure}
\center
\includegraphics[totalheight=8.5cm,width=8.5cm]{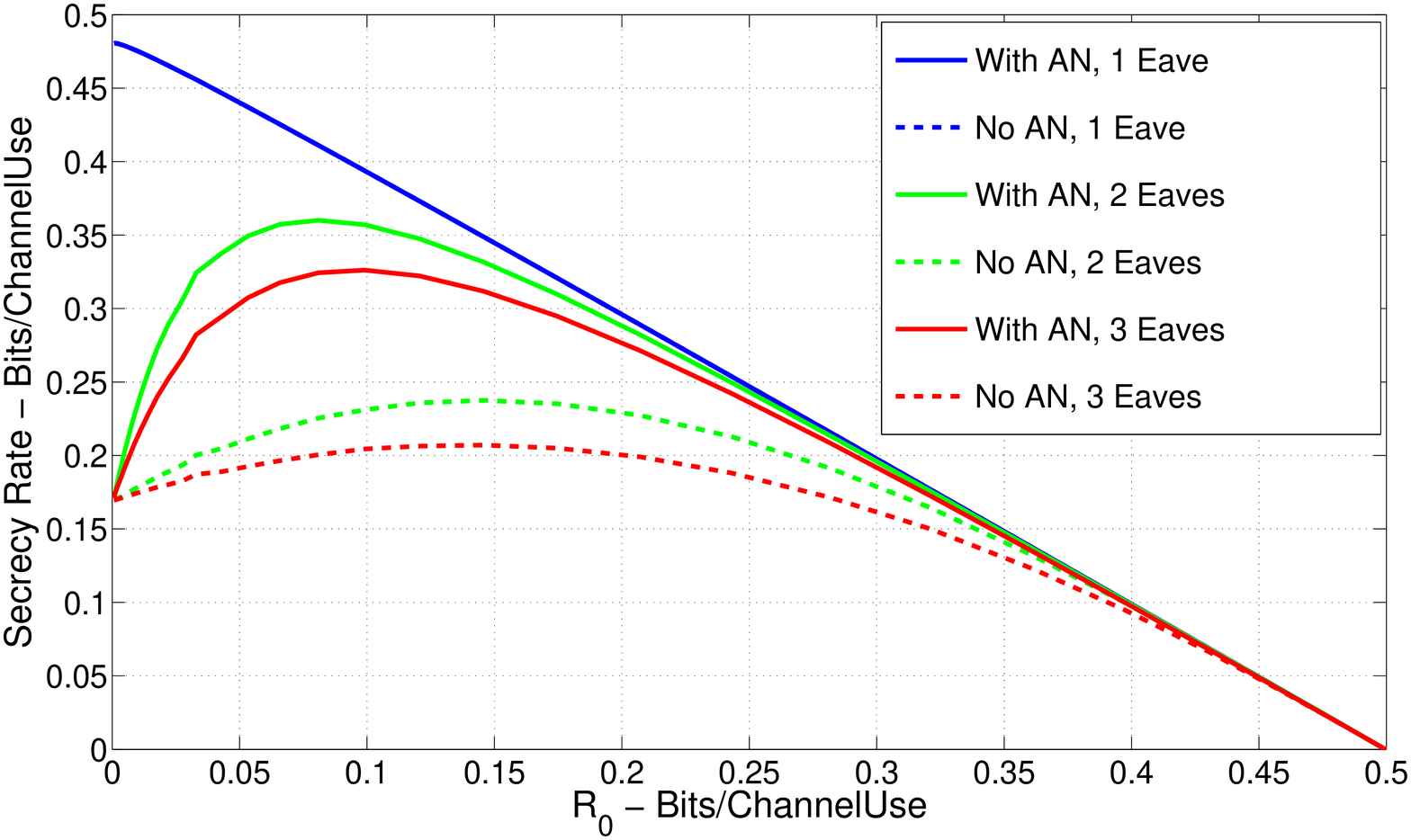}
\caption{Secrecy rate vs $R_{0}$ in MIMO DF relay beamforming for BPSK 
alphabet, with/without AN signal. $N=2$, $N_0=1$, $J=1,2,3$, $M=2$, 
fixed $P_s = 0$ dB, and $P_R = 9$ dB.}
\label{sim_fig_mimo_df_relay_bf_jamming_finitealphabet_1}
\end{figure}

{\em Perfect CSI case of Section 
\ref{sec_mimo_df_relay_bf_jamming_finitealphabet_3} :}
We have used the following channel gains in the simulations:
\begin{eqnarray}
\boldsymbol{g}      &=& [-0.5839 + 2.2907i,  -0.7158 + 0.1144i]^{T}, 
\label{eqn_mimo_df_relay_bf_jamming_finitealphabet_66} \\
h_0                 &=& -0.3822 - 0.3976i, 
\label{eqn_mimo_df_relay_bf_jamming_finitealphabet_67} \\
z_{01}              &=&  0.0123 + 0.0137i,                           
\label{eqn_mimo_df_relay_bf_jamming_finitealphabet_68} \\
z_{02}              &=&  0.0231 - 0.0178i,                           
\label{eqn_mimo_df_relay_bf_jamming_finitealphabet_69} \\
z_{03}              &=& -0.0045 - 0.0042i,                           
\label{eqn_mimo_df_relay_bf_jamming_finitealphabet_70} \\
\boldsymbol{h}      &=& [0.2174 - 0.6913i, \ -0.4047 - 0.3159i],     
\label{eqn_mimo_df_relay_bf_jamming_finitealphabet_71} \\
\boldsymbol{z}_{1}  &=& [0.3826 + 0.0811i, \ 0.8389 - 0.0943i],      
\label{eqn_mimo_df_relay_bf_jamming_finitealphabet_72} \\
\boldsymbol{z}_{2}  &=& [0.2977 + 0.7902i, \ -0.2069 + 0.4696i],     
\label{eqn_mimo_df_relay_bf_jamming_finitealphabet_73} \\
\boldsymbol{z}_{3}  &=& [-0.6076 + 0.6637i, \ -0.3316 + 0.1921i].    
\label{eqn_mimo_df_relay_bf_jamming_finitealphabet_74}
\end{eqnarray}
In Fig. \ref{sim_fig_mimo_df_relay_bf_jamming_finitealphabet_1}, 
we plot the secrecy rate versus $R_{0}$ for BPSK alphabet 
(i.e., $M =2$), with/without AN, $J = 1,2,3$ eavesdroppers, 
$P_{s} = 0$ dB, and $P_{R} = 9$ dB. We observe that the secrecy rate 
initially increases with increase in $R_{0}$ and then drops to zero
for large values of $R_{0}$. We also observe that the injection of AN 
improves the secrecy rate when $J= 2,3$ eavesdroppers are present. 
However, when only one eavesdropper is present, the secrecy rate plots 
with/without AN overlap. This is due to the null signal beamforming by 
the MIMO relay at the eavesdropper. This is possible only when the number 
of eavesdroppers is strictly less than the number of antennas in the 
MIMO relay which happens to be true for this case with $N=2$ and $J=1$. 
For the case when $J=1$, the secrecy rate maximum happens at
$R_{0}=0.001445$. Further, for the case when $J=2,3$ and without AN, the 
secrecy rate maximum happens at $R_{0}=0.145797$, and with AN it happens 
at $R_{0}=0.080959$ and $R_{0}=0.099059$, respectively. 
It is seen that the secrecy rate falls 
approximately linearly for large $R_{0}$.
The near-linear fall in secrecy rate for large values of $R_{0}$
is due to the saturation of $S-D$ link information rate to $\frac{1}{2}\log_{2} 2 = 0.5$
for $M = 2$. We have also numerically 
observed that the rank of $\boldsymbol{\Phi}$ is 1.  

\begin{figure}[t]
\center
\includegraphics[totalheight=8.5cm,width=8.5cm]{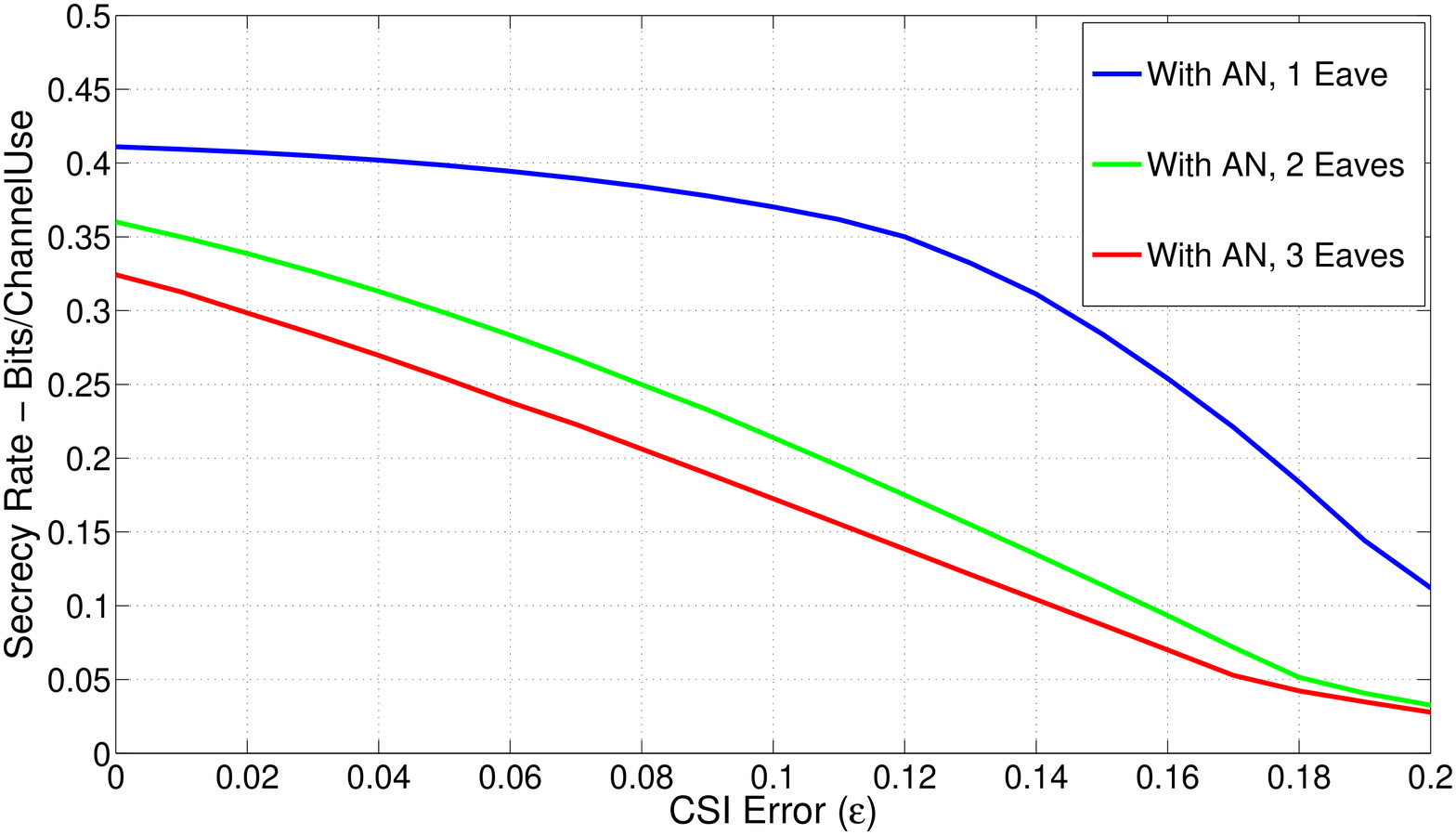}
\caption{ $R_{s}$ vs $\epsilon$ in MIMO DF relay beamforming for BPSK 
alphabet and with AN signal. $N=2$, $N_0=1$, $J=1,2,3$, $M=2$, fixed 
$R_{0} = 0.0810$, $P_{s} = 0$ dB, and $P_{R} = 9$ dB.} 
\label{sim_fig_mimo_df_relay_bf_jamming_finitealphabet_2}
\end{figure}

{\em Imperfect CSI case of Section 
\ref{sec_mimo_df_relay_bf_jamming_finitealphabet_4} :}
Here, we assume that the channel gains in 
(\ref{eqn_mimo_df_relay_bf_jamming_finitealphabet_66})-(\ref{eqn_mimo_df_relay_bf_jamming_finitealphabet_74}) 
are the available CSI estimates. We also assume that the magnitudes 
of the CSI errors in all the links are equal, i.e., 
$\epsilon_{ \boldsymbol{g} } = \epsilon_{ h_{0} } = \epsilon_{ z_{0j} } = \epsilon_{ \boldsymbol{h} } = \epsilon_{ \boldsymbol{z}_{j} } = \epsilon$.
We solve the optimization problem 
(\ref{eqn_mimo_df_relay_bf_jamming_finitealphabet_55}) for BPSK alphabet 
(i.e., $M =2$), with AN, fixed $R_{0} = 0.0810$, $P_{s} = 0$ dB, and 
$P_{R} = 9$ dB. In Fig. 
\ref{sim_fig_mimo_df_relay_bf_jamming_finitealphabet_2}, 
we plot $R_{s}$ vs $\epsilon$ with AN for $J = 1,2,3$.
We observe that the secrecy rate decreases with increase in CSI error 
and with increase in number of eavesdroppers. We have also numerically 
observed that the rank of $\boldsymbol{\Phi}$ is 1.

\section{Conclusions}
\label{sec_mimo_df_relay_bf_jamming_finitealphabet_6}
We considered MIMO DF relay beamforming with imperfect CSI, cooperative 
artificial noise injection, and finite-alphabet input in the presence of 
an user and multiple non-colluding eavesdroppers. The source transmits 
common and secret messages which are intended for the user. The common
 message is transmitted at a fixed rate $R_{0}$. In order to maximize the 
worst case secrecy rate, we maximized the worst case link information
rate to the user subject to: $i)$ the individual power constraints on 
the source and the MIMO DF relay, and $ii)$ the best case link information 
rates to $J$ eavesdroppers be less than or equal to $R_{0}$ in order to 
support a fixed common message rate $R_{0}$. Numerical results showing the 
effect of perfect/imperfect CSI, presence/absence of AN with finite-alphabet 
input on the secrecy rate were presented. We would like to remark that the 
work presented in this paper can be extended to amplify-and-forward relay 
channel.

\section*{Appendix} 
In this appendix, we analyze the rank of the optimal solution 
$\boldsymbol{\Phi}$ obtained from the optimization problem 
(\ref{eqn_mimo_df_relay_bf_jamming_finitealphabet_25}). We take the 
Lagrangian of the objective function $-t$ with constraints in 
(\ref{eqn_mimo_df_relay_bf_jamming_finitealphabet_59})-(\ref{eqn_mimo_df_relay_bf_jamming_finitealphabet_26}) as follows \cite{ir25}:
\begin{eqnarray}
\ell(t, \ \boldsymbol{\Phi}, \ \boldsymbol{\Psi}, \ \lambda, \ \boldsymbol{\Lambda}_{1}, \ \boldsymbol{\Lambda}_{2}, \ \mu, \ \nu_{j}, \ \xi) \ = 
\ - t - \Tr(\boldsymbol{\Lambda}_{1}\boldsymbol{\Phi}) \nonumber \\ - \Tr(\boldsymbol{\Lambda}_{2}\boldsymbol{\Psi}) 
+ \lambda\Big( \Tr(\boldsymbol{\Phi}) + \Tr(\boldsymbol{\Psi}) - P_{R} \Big) \nonumber \\
+ \mu \Big( (t - a) \big( N_{0} + \boldsymbol{h} \boldsymbol{\Psi} \boldsymbol{h}^{\ast} \big) - (\boldsymbol{h} \boldsymbol{\Phi} \boldsymbol{h}^{\ast}) \Big) \nonumber \\ 
+ \sum^{J}_{j = 1} \nu_{j}\Big( \big( \boldsymbol{z}_{j} \boldsymbol{\Phi} \boldsymbol{z}^{\ast}_{j} \big) - \Big( I^{-1}(2R_{0}) - b_{j} \Big) ( N_{0} + \boldsymbol{z}_{j} \boldsymbol{\Psi} \boldsymbol{z}^{\ast}_{j}  ) \Big) \nonumber \\ 
+ \xi \Big( (\boldsymbol{h} \boldsymbol{\Phi} \boldsymbol{h}^{\ast}) - (c - a) \big( N_{0} + \boldsymbol{h} \boldsymbol{\Psi} \boldsymbol{h}^{\ast} \big) \Big)
\label{eqn_mimo_df_relay_bf_jamming_finitealphabet_58}
\end{eqnarray}
where 
$\lambda \geq 0, \ \boldsymbol{\Lambda}_{1} \succeq \boldsymbol{0}, \ \boldsymbol{\Lambda}_{2} \succeq \boldsymbol{0}, \ \mu \geq 0, \ \nu_{j} \geq 0, \ \xi \geq 0$ 
are Lagrangian multipliers. The KKT conditions for 
(\ref{eqn_mimo_df_relay_bf_jamming_finitealphabet_58}) are as follows:
\begin{itemize}
\vspace{2mm}
\item[$(a1)$] all constraints in 
(\ref{eqn_mimo_df_relay_bf_jamming_finitealphabet_59})-(\ref{eqn_mimo_df_relay_bf_jamming_finitealphabet_26}), 
\vspace{2mm}
\item[$(a2)$] $\Tr(\boldsymbol{\Lambda}_{1}\boldsymbol{\Phi}) = 0$. Since $\boldsymbol{\Lambda}_{1} \succeq \boldsymbol{0}$ and 
$\boldsymbol{\Phi} \succeq \boldsymbol{0}$ $\implies$ $\boldsymbol{\Lambda}_{1}\boldsymbol{\Phi} = \boldsymbol{0}$,
\vspace{2mm}
\item[$(a3)$] $\Tr(\boldsymbol{\Lambda}_{2}\boldsymbol{\Psi}) = 0$. Since $\boldsymbol{\Lambda}_{2} \succeq \boldsymbol{0}$ and 
$\boldsymbol{\Psi} \succeq \boldsymbol{0}$ $\implies$ $\boldsymbol{\Lambda}_{2}\boldsymbol{\Psi} = \boldsymbol{0}$,
\vspace{2mm}
\item[$(a4)$] $\lambda \Big( \Tr(\boldsymbol{\Phi}) + \Tr(\boldsymbol{\Psi}) - P_{R} \Big) \ = \ 0$,
\vspace{2mm}
\item[$(a5)$] $ \mu \Big( (t - a) \big( N_{0} + \boldsymbol{h} \boldsymbol{\Psi} \boldsymbol{h}^{\ast} \big) - (\boldsymbol{h} \boldsymbol{\Phi} \boldsymbol{h}^{\ast}) \Big) \ = \ 0$,
\vspace{2mm}
\item[$(a6)$] $ \forall j =1,2,\cdots,J, \ \ \nu_{j}\Big( \big( \boldsymbol{z}_{j} \boldsymbol{\Phi} \boldsymbol{z}^{\ast}_{j} \big) - \Big( I^{-1}(2R_{0}) - b_{j} \Big) ( N_{0} + \boldsymbol{z}_{j} \boldsymbol{\Psi} \boldsymbol{z}^{\ast}_{j}  ) \Big) \  = \ 0$,
\vspace{2mm}
\item[$(a7)$] $ \xi \Big( (\boldsymbol{h} \boldsymbol{\Phi} \boldsymbol{h}^{\ast}) - (c - a) \big( N_{0} + \boldsymbol{h} \boldsymbol{\Psi} \boldsymbol{h}^{\ast} \big) \Big) \ = \ 0$,
\vspace{2mm}
\item[$(a8)$] $\frac{\partial \ell}{\partial t} = 0$ $\implies$ $\mu \big( N_{0} + \boldsymbol{h} \boldsymbol{\Psi} \boldsymbol{h}^{\ast} \big) = 1$. This further implies that $\mu > 0$,
\vspace{2mm}
\item[$(a9)$] $\frac{\partial \ell}{\partial \boldsymbol{\Phi}} = \boldsymbol{0}$ $\implies$ $\boldsymbol{\Lambda}_{1} = \lambda \boldsymbol{I} - \mu ( \boldsymbol{h}^{\ast} \boldsymbol{h} ) + \sum^{J}_{j=1} \nu_{j} ( \boldsymbol{z}^{\ast}_{j} \boldsymbol{z}_{j} ) + \xi ( \boldsymbol{h}^{\ast} \boldsymbol{h} )$,
\vspace{2mm}
\item[$(a10)$] $\frac{\partial \ell}{\partial \boldsymbol{\Psi}} = \boldsymbol{0}$ $\implies$ $\boldsymbol{\Lambda}_{2} = \lambda \boldsymbol{I} + \mu (t - a)( \boldsymbol{h}^{\ast} \boldsymbol{h} ) - \sum^{J}_{j=1} \nu_{j}\Big( I^{-1}(2R_{0}) - b_{j} \Big)( \boldsymbol{z}^{\ast}_{j} \boldsymbol{z}_{j} ) - \xi (c-a)( \boldsymbol{h}^{\ast} \boldsymbol{h} ) $.
\end{itemize}
The KKT conditions $(a8)$ and $(a5)$ imply that the constraint
(\ref{eqn_mimo_df_relay_bf_jamming_finitealphabet_59}) will be 
satisfied with equality. Assuming $\boldsymbol{\Phi} \neq \boldsymbol{0}$, 
this further implies that $t > a$. The constraints 
(\ref{eqn_mimo_df_relay_bf_jamming_finitealphabet_60}) and 
(\ref{eqn_mimo_df_relay_bf_jamming_finitealphabet_61}) imply that 
$I^{-1}(2R_{0}) \ge b_{j}$ and $c > a$. The KKT conditions 
$(a9)$, $(a10)$, $(a2)$, $(a3)$, $(a4)$, $(a5)$, $(a6)$, and $(a7)$
imply that
\begin{eqnarray}
\lambda P_{R} -\mu (t-a)N_{0} + \sum^{J}_{j = 1} \nu_{j} \Big( I^{-1}(2R_{0}) - b_{j} \Big)N_{0} \nonumber \\ + \xi (c-a)N_{0} = 0.
\label{eqn_mimo_df_relay_bf_jamming_finitealphabet_62} 
\end{eqnarray}
Let $P_{R}$ be small enough 
such that the constraint in
(\ref{eqn_mimo_df_relay_bf_jamming_finitealphabet_61})
is satisfied with strict inequality. This implies that the KKT condition
$(a7)$ will be satisfied only when $\xi = 0$.
With $\xi = 0$, we consider the scenario when the expression 
(\ref{eqn_mimo_df_relay_bf_jamming_finitealphabet_62}) is satisfied 
for $\lambda > 0$. With $\lambda > 0$, the KKT condition $(a4)$ 
implies that $\Tr(\boldsymbol{\Phi}) + \Tr(\boldsymbol{\Psi}) = P_{R}$, 
i.e., the entire relay power, $P_{R}$, will be used for transmission.
Further, we rewrite the KKT condition $(a9)$ as follows:
\begin{eqnarray}
\boldsymbol{\Lambda}_{1} + \mu ( \boldsymbol{h}^{\ast} \boldsymbol{h} ) \ = \ \lambda \boldsymbol{I} + \sum^{J}_{j=1} \nu_{j} ( \boldsymbol{z}^{\ast}_{j} \boldsymbol{z}_{j} ) \succ \boldsymbol{0}.
\label{eqn_mimo_df_relay_bf_jamming_finitealphabet_63} 
\end{eqnarray}
The above expression implies that 
$rank\big( \boldsymbol{\Lambda}_{1} ) \geq N - rank \big(\mu ( \boldsymbol{h}^{\ast} \boldsymbol{h} ) \big) = N-1$.
The KKT condition $(a2)$ further implies that 
$rank\big( \boldsymbol{\Lambda}_{1})=N-1$ and $rank\big( \boldsymbol{\Phi})=1$.

We now show that the solution $\boldsymbol{\Phi}$ of the optimization 
problem (\ref{eqn_mimo_df_relay_bf_jamming_finitealphabet_25}) has rank-1 
even for large values of $P_{R}$. Let 
$\boldsymbol{\Phi} \neq  \boldsymbol{0} \ (\succeq \boldsymbol{0})$ and 
$\boldsymbol{\Psi} \neq  \boldsymbol{0} \ (\succeq \boldsymbol{0})$ be the optimal solutions of 
(\ref{eqn_mimo_df_relay_bf_jamming_finitealphabet_25}) with
\begin{eqnarray}
\Tr(\boldsymbol{\Phi})+\Tr(\boldsymbol{\Psi}) \ = \ P \ \leq \ P_{R}, \nonumber
\end{eqnarray}
and the objective function value $t > 0$. Define 
\begin{eqnarray}
\boldsymbol{\Phi}_{0} \ &=& \ \frac{\boldsymbol{\Phi}}{\Tr(\boldsymbol{\Phi}) + \Tr(\boldsymbol{\Psi})} \ = \ \frac{\boldsymbol{\Phi}}{P}, 
\nonumber \\
\boldsymbol{\Psi}_{0} \ &=& \ \frac{\boldsymbol{\Psi}}{\Tr(\boldsymbol{\Phi}) + \Tr(\boldsymbol{\Psi})} \ = \ \frac{\boldsymbol{\Psi}}{P}. \nonumber
\end{eqnarray}
It is obvious that the objective function value, $t$, in the 
optimization problem (\ref{eqn_mimo_df_relay_bf_jamming_finitealphabet_25}) 
is a non-decreasing function in $P_{R}$. As discussed previously for small 
values of $P_{R}$, the optimization problem 
(\ref{eqn_mimo_df_relay_bf_jamming_finitealphabet_25}) 
attains it's maximum value when entire power is used, i.e., 
$(\boldsymbol{\Phi}_{}, \ \boldsymbol{\Psi}_{}) = (P_{}\boldsymbol{\Phi}_{0}, \ P_{}\boldsymbol{\Psi}_{0}) = (P_{R}\boldsymbol{\Phi}_{0}, \ P_{R}\boldsymbol{\Psi}_{0})$. 
This implies that the objective function value, $t$, in 
(\ref{eqn_mimo_df_relay_bf_jamming_finitealphabet_25}) is a 
strictly increasing function in $P_{R}$ for small values of $P_{R}$.
We now fix the directional matrices 
$(\boldsymbol{\Phi}_{0}, \ \boldsymbol{\Psi}_{0})$
which are obtained for small values of $P_{R}$ such that the 
constraint in (\ref{eqn_mimo_df_relay_bf_jamming_finitealphabet_61}) is satisfied 
with strict inequality. We rewrite the constraints in
(\ref{eqn_mimo_df_relay_bf_jamming_finitealphabet_60}) and 
(\ref{eqn_mimo_df_relay_bf_jamming_finitealphabet_61}) in the following 
forms, respectively:

\vspace{-2mm}
{\small
\begin{eqnarray}
\forall j=1,2,\cdots,J, \quad I^{-1}(2R_{0}) \ \geq \ \bigg( b_{j} +  \frac{ \boldsymbol{z}_{j} \boldsymbol{\Phi} \boldsymbol{z}^{\ast}_{j} }{ N_{0} + \boldsymbol{z}_{j} \boldsymbol{\Psi} \boldsymbol{z}^{\ast}_{j} } \bigg) , 
\label{eqn_mimo_df_relay_bf_jamming_finitealphabet_64} \\
c \ \geq \ \bigg( a + \frac{ \boldsymbol{h} \boldsymbol{\Phi} \boldsymbol{h}^{\ast} }{ N_{0} + \boldsymbol{h} \boldsymbol{\Psi} \boldsymbol{h}^{\ast}  } \bigg) 
\label{eqn_mimo_df_relay_bf_jamming_finitealphabet_65}.
\end{eqnarray}
}

\vspace{-2mm} 
\hspace{-3.5mm}
In the above inequalities, the derivatives of the functions
$(\frac{ \boldsymbol{z}_{j} \boldsymbol{\Phi} \boldsymbol{z}^{\ast}_{j} }{ N_{0} + \boldsymbol{z}_{j} \boldsymbol{\Psi} \boldsymbol{z}^{\ast}_{j} } )$ and 
$( \frac{\boldsymbol{h}^{}\boldsymbol{\Phi}\boldsymbol{h}^{\ast}}{N_{0} + \boldsymbol{h}^{} \boldsymbol{\Psi} \boldsymbol{h}^{\ast}} )$
w.r.t. $P$ when evaluated at 
$(P\boldsymbol{\Phi}_{0}, \ P\boldsymbol{\Psi}_{0})$ 
are $\geq 0$ and $>0$, respectively. This implies that the right hand 
sides of the inequalities in 
(\ref{eqn_mimo_df_relay_bf_jamming_finitealphabet_64}) and 
(\ref{eqn_mimo_df_relay_bf_jamming_finitealphabet_65}) are 
non-decreasing and strictly increasing functions in $P$, respectively, at 
$(P\boldsymbol{\Phi}_{0}, \ P\boldsymbol{\Psi}_{0})$. 
This further implies that if the above inequalities are satisfied at 
$(P_{R}\boldsymbol{\Phi}_{0}, \ P_{R}\boldsymbol{\Psi}_{0})$, the 
optimization problem (\ref{eqn_mimo_df_relay_bf_jamming_finitealphabet_25}) 
will attain its maximum value at 
$(P_{R}\boldsymbol{\Phi}_{0}, \ P_{R}\boldsymbol{\Psi}_{0})$, 
i.e., when the entire available relay power, $P_{R}$, is used.
When $P_{R}$ is large such that the above inequalities fail to satisfy 
at $(P_{R}\boldsymbol{\Phi}_{0}, \ P_{R}\boldsymbol{\Psi}_{0})$,
the optimization problem 
(\ref{eqn_mimo_df_relay_bf_jamming_finitealphabet_25}) will attain its 
maximum value at $(P\boldsymbol{\Phi}_{0}, \ P\boldsymbol{\Psi}_{0})$, 
where $P \ (< P_{R})$ is the maximum power at which the above inequalities 
are satisfied at $(P\boldsymbol{\Phi}_{0}, \ P\boldsymbol{\Psi}_{0})$.
The excess power $(P_{R} - P)$ will remain unused. This implies that the 
ranks of $\boldsymbol{\Phi}$ and $\boldsymbol{\Psi}$ remain constant for 
large values of $P_{R}$.

\end{document}